\documentclass[12pt,a4paper]{article}
\usepackage[T1]{fontenc}
\usepackage[utf8]{inputenc}
\usepackage{authblk}
\usepackage{amsmath,amssymb,epsfig,cite,dsfont}
\usepackage{array,multirow}
\usepackage {color}
\usepackage[toc,page]{appendix}
\numberwithin{equation}{section}

\usepackage[colorlinks=true
,urlcolor=blue
,anchorcolor=blue
,citecolor=blue
,filecolor=blue
,linkcolor=blue
,menucolor=blue
,pagecolor=blue
,linktocpage=true
,pdfproducer=medialab
,pdfa=true
]{hyperref}

\advance\hoffset by -1.1truecm
\advance\voffset by -1.0truecm
\newcommand{\be}{\begin{equation}}
\newcommand{\ee}{\end{equation}}
\newcommand{\bea}{\begin{eqnarray}}
\newcommand{\eea}{\end{eqnarray}}

\usepackage{mathrsfs}
\usepackage{authblk}

\numberwithin{equation}{section}
\usepackage{float}
\restylefloat{figure}
\newcounter{appendice}

\begin{document}

 \title{
 \begin{flushright}
 \end{flushright}
\vspace{2.2cm}\begin{flushleft} \bf { Monopoles, Dirac operator and index theory for fuzzy  ${SU(3)}/({U(1)\times U(1)})$
\linethickness{.05cm}\line(1,0){433}
}\end{flushleft}}
\author[a]{\bf Nirmalendu~Acharyya\footnote{nirmalendu@cts.iisc.ernet.in}}
\author[a,b]{\bf Ver\'onica~Errasti~D\'iez\footnote{vediez@physics.mcgill.ca}}
\affil[a]{ \small Centre for High Energy Physics,  Indian Institute of Science, Bangalore-560012,
India}
\affil[b]{ \hspace{0cm} Physics Department, McGill University, 3600 University St., Montreal, QC H3A 2T8, Canada}
\date{\empty}
\maketitle
\abstract{The intersection of the 10-dimensional fuzzy conifold $Y_F^{10}$ with $S^5_F \times S^5_F$ is the compact 8-dimensional fuzzy space $X_F^8$. 
We show that $X_F^8$ is (the analogue of) a principal $U(1)\times U(1)$ bundle over fuzzy ${SU(3)}/({U(1) \times U(1)})  \left(\equiv\mathcal{M}^6_F\right)$.
We construct $\mathcal{M}_F^6$ using the Gell-Mann matrices by adapting Schwinger's  construction. The space $\mathcal{M}_F^6$ is of relevance in higher dimensional quantum Hall effect and  matrix models of $D$-branes.

Further we show that the sections of the monopole bundle can be expressed in the basis of $SU(3)$ eigenvectors. We construct the Dirac operator on $\mathcal{M}_F^6$ from the Ginsparg-Wilson algebra on this space. Finally, we show that the index of the Dirac operator correctly reproduces the known results in the continuum.   }

\section{Introduction}

Fuzzy spaces emerge naturally in the discussion of various theories like   quantum Hall effect (QHE) and matrix models of $D$-branes in the presence of certain background fields.
%
%
%
%
%
%
%
%
%
%
In the context of QHE in 2-dimensions, the Hilbert space of the lowest Landau level corresponds to symmetric representations of $SU(2)$. The observables for the lowest Landau level then correspond to observables of $S^2_F$ (for example, see \cite{Susskind:2001fb, Polychronakos:2001mi, Karabali:2004xq}). 

Higher dimensional $(d>2)$ generalizations of QHE are interesting for various reasons. For example, they extend the notion of incompressibility to higher dimensions.  In QHE in $d >2$, the Landau problem is replaced by a particle moving on a compact coset space  in the presence of background gauge fields (say, monopoles). One such coset space is $SU(3)/(U(1) \times U(1))$.  The intersection of the 10-dimensional conifold $Y^{10}$ $(\equiv \{z_\alpha, w_\beta:\,\, z_\alpha w_\alpha =0, \,\, \alpha,\beta=1,2,3\})$  with $S^5\times S^5$ is a $U(1) \times U(1)$ monopole bundle on this coset space. For a particle moving on this coset space in the presence of background $U(1)$ monopoles, the Hilbert space of  the lowest Landau level has an exact correspondence with the representations of $SU(3)$ (for instance, see \cite{Karabali:2006eg}).
Consequently, the observables for the lowest Landau level are observables of fuzzy $SU(3)/(U(1)\times U(1))$$(\equiv\mathcal{M}_F^6)$. Thus in the presence of background fields, the natural description of the space becomes fuzzy and the emergent compact fuzzy space like $\mathcal{M}_F^6$ and monopoles on it are relevant in the understanding of such QHE.

$\mathcal{M}_F^6$ is described by  matrix algebras on the carrier space of the representations of $SU(3)$. This space appears in the study of matrix models  describing branes. The low energy effective action of a $N$ (coincident) $D$-brane system is that of a $U(N)$ Yang-Mills theory. It is well known that the corresponding  transverse geometry is inherently noncommutative  \cite{Basu:2004ed}. Owing to the non-Abelian nature of such theories, the $N$ $Dp$-brane system couples to Ramond-Ramond field strengths of degree $\geq p+ 4$  \cite{Myers:1999ps, Trivedi:2000mq}. In particular, when the RR background is chosen to be proportional to the $SU(3)$ structure constants,  an 8-matrix model has  the action
\begin{eqnarray}\label{matrix_action}
S = T_0 Tr\left[\frac{1}{2}\dot{ \phi_i}^2+\frac{1}{4} [\phi_i, \phi_j]^2 -\frac{i}{3}\kappa f_{ijk} \phi_i[\phi_j, \phi_k]\right], \quad\quad i=1,2,\ldots, 8.
\end{eqnarray}
Here, $\kappa$ is a coupling constant, $\phi_i$'s  are $N\times N$ matrices and $f_{ijk}$'s are the $SU(3)$ structure constants.
This describes $N$ coincident $Dp$-branes (with $ p \leq 1$). One of the ground state configurations of the above action is 
$\mathcal{M}_F^6$  \cite{Kimura:2003kd}.

${SU(3)}/({U(1) \times U(1)})$ is of particular relevance in string theory. For instance, a 7-dimensional  space with $G_2$ holonomy can have a conical singularity on  this space  \cite{Atiyah:2001qf}. Understanding how  this holonomy appears in the fuzzy case is an interesting question by itself. We leave it for a seperate investigation in  future. Here we focus primarily on the construction of $\mathcal{M}_F^6$.

To discuss higher dimensional QHE , we should introduce fermions (electrons) on $\mathcal{M}_F^6$.
 To this end, we construct the Dirac operator on $\mathcal{M}_F^6$.   The Dirac operator is also necessary to discuss supersymmetry on this fuzzy space, which is  of interest to many. In a seperate context,  since this is a finite dimensional model, it is important to study the fermion-doubling problem with the Dirac operator.
The construction of the Dirac operator on a fuzzy coset space like $\mathcal{M}_F^6$ is nontrivial and intrinsically interesting.
A beautiful (expected) relation exists between the index of the Dirac operator and the topological objects on $\mathcal{M}_F^6$, which we make explicit.

Our construction of  $\mathcal{M}_F^6$ derives from a Schwinger-like construction using the Gell-Mann matrices and  six independent oscillators. With these six oscillators, the  10-dimensional fuzzy conifold $Y^{10}_F$ can be constructed  as in \cite{Acharyya:2013hga, Acharyya:2013oma}. $X^8_F=Y^{10}_F\cap\left( S^5_F \times S^5_F\right)$ describes a subspace of this 6-dimensional oscillator's Hilbert space, which  is the carrier space of all the representations of $SU(3)$. In section \ref{sec_2}, we show that there exists a Hopf-like map $X^8_F \rightarrow\mathcal{M}_F^6$ .

 $X^8_F$ is a  $U(1) \times U(1)$ principal bundle over $\mathcal{M}_F^6$. The monopoles can be characterized by linear maps from one representation space of $SU(3)$ to another \cite{Acharyya:2013hga, Acharyya:2013oma, Grosse:1995jt, Grosse:1996mz, Acharyya:2014nfa}.  Such sections are  rectangular matrices that map
 a  $\mathcal{M}_F^6$ of a given size to another of different size. In section \ref{sec_3}, we construct these matrices in the basis of the $SU(3)$ $D$-matrices.


In section \ref{sec_4}, we show that a  Ginsparg-Wilson (GW) algebra is associated with $\mathcal{M}^6_F$. The Dirac operator can be constructed using the elements of this GW algebra, which are functions of the generators of $SU(3)$.  We compute the index of the Dirac operator using the representations of $SU(3)$ and their quadratic Casimir values as in    \cite{Dolan:2003bj}. The index  is equal to $\mathrm{Tr}\, (F\wedge F\wedge F)$.


\section{Construction of fuzzy ${SU(3)}/({U(1)\times U(1)})$ }\label{sec_2}
 $\mathbb{C}^6_F$ is described by six independent oscillators $\hat{a}_\alpha, \hat{b}_\alpha$ ($\alpha=1,2,3$):
\begin{eqnarray}
\begin{array}{lll}
& \Big[ \hat{a}_\alpha, \hat{a}_\beta\Big]=0, & \quad  \left[ \hat{a}_\alpha, \hat{a}_\beta^\dagger\right]=\delta_{\alpha\beta},   \\ \\
& \left[ \hat{b}_\alpha, \hat{b}_\beta\right]=0,& \quad \left[ \hat{b}_\alpha, \hat{b}_\beta^\dagger\right]=\delta_{\alpha\beta}, \\ \\
& \left[ \hat{a}_\alpha, \hat{b}_\beta\right]=0, & \quad \left[ \hat{a}_\alpha, \hat{b}_\beta^\dagger\right]=0.
\end{array}
\end{eqnarray}
These oscillators act on the Hilbert space $\mathcal{F}$ spanned by the eigenstates of the number operators $\hat{N}_a \,\,(\equiv \hat{a}^\dagger_\alpha \hat{a}_\alpha)$ and $\hat{N}_b \,\,(\equiv \hat{b}^\dagger_\alpha\hat{b}_\alpha)$:
\begin{eqnarray}
\mathcal{F} \equiv span \left\{ |n_a^1, n_a^2, n_a^3; n_b^1 n_b^2, n_b^3\rangle: n_a^\alpha, n_b^\alpha=0,1,2,\ldots  \right\}. 
\end{eqnarray}

A fuzzy conifold is described by a subalgebra in $\mathbb{C}^6_F$ \cite{Acharyya:2013hga, Acharyya:2013oma}.
We  define  the operator 
\begin{eqnarray}
\hat{\mathcal{O}} \equiv \sum_{\alpha=1}^3\hat{b}_\alpha\hat{a}_\alpha,
\end{eqnarray}
which has as its kernel 
\begin{eqnarray}
\ker( \hat{\mathcal{O}}) =  span \left\{ |\cdot\rangle\in \mathcal{F}: \,\,\, \hat{\mathcal{O}}  |\cdot\rangle=0     \right \} \subset \mathcal{F} .
\end{eqnarray}
(We use the symbol $| \cdot \rangle$  to denote the state $|n_a^1, n_a^2, n_a^3; n_b^1 n_b^2, n_b^3\rangle$.)
The algebra of $(\hat{a}_\alpha, \hat{b}_\beta)$'s restricted to $\ker( \hat{\mathcal{O}})$ describes a 10-dimensional fuzzy conifold $Y_F^{10}$.

For convenience of normalization, we will also work with the set of operators 
\begin{equation}
\hat{\chi}_\alpha\equiv\hat{a}_\alpha \frac{1}{\sqrt{N_a}}, \quad \quad \hat{\xi}_\alpha\equiv\hat{b}_\alpha \frac{1}{\sqrt{N_b}},
\end{equation}
which satisfy
\begin{equation}
\hat{\chi}_\alpha^\dagger\hat{\chi}_\alpha=1, \quad \quad \hat{\xi}_\alpha^\dagger\hat{\xi}_\alpha=1.
\end{equation}
The $\hat{\chi}_\alpha$'s (or $\hat{\xi}_\alpha$'s) are well-defined if we exclude the states for which $\hat{N}_a=0$ (or $\hat{N}_b=0$).
Then the algebra generated by $\hat{\chi}_\alpha$'s and $\hat{\xi}_\alpha$'s  describes $S_F^5\times S_F^5$.

The operator $\hat{\mathcal{O}}^\prime\equiv \hat{\xi}_\alpha \hat{\chi}_\alpha$ also vanishes in $\ker (\hat{\mathcal{O}})$. Therefore the algebra of the
$\hat{\chi}_\alpha$'s and $\hat{\xi}_\alpha$'s restricted to $\ker (\hat{\mathcal{O}})$ describes an 8-dimensional fuzzy space $X_F^8$.
Informally, we can think of $X^8_F$  as $Y_F^{10}\cap \left(S_F^5\times S_F^5\right)$.

\subsection{$X^8_F \rightarrow\mathcal{M}_F^6\equiv$  fuzzy ${SU(3)}/({U(1)\times U(1)})$}
Using the matrices $
T_i=\frac{1}{2}\lambda_i$  ($
\lambda_i=$ Gell-Mann matrices, $ i=1,2,\ldots,8$) satisfying
\begin{eqnarray}
[T_i,T_j]=if_{ijk}T_k, \quad \quad  \{T_i,T_j\}=\frac{1}{3}\delta_{ij}+d_{ijk}T_k,
\end{eqnarray}
we can define a Schwinger-like construction (similar to  \cite{Mathur:2000sv}):
\begin{eqnarray}
\label{map1}
&&\hat{y}_i =  \hat{a}_\alpha^\dagger (T_i)^{\alpha\beta} \hat{a}_\beta -  \hat{b}_\alpha (T_i)^{\alpha\beta} \hat{b}_\beta^\dagger, \\
&&\hat{s}_i= \hat{\chi}_\alpha^\dagger (T_i)^{\alpha\beta} \hat{\chi}_\beta- \hat{\xi}_\alpha (T_i)^{\alpha\beta} \hat{\xi}_\beta^\dagger. 
\end{eqnarray}
The $\hat{y}_i$'s obey
\begin{equation}\label{algebra_12}
 \hat{y}_i^\dagger=\hat{y}_i, \quad\quad  [\hat{y}_i,\hat{y}_j]=if_{ijk}\hat{y}_k
\end{equation}
and the  Casimirs are $ \hat{C}_2\equiv \hat{y}_i\hat{y}_i$ and $\hat{C}_3\equiv d_{ijk}\hat{y}_i\hat{y}_j\hat{y}_k$:
\begin{eqnarray}\label{quadratic_casimir_12}
&&\hspace*{-.9cm}\hat{C}_2 = \frac{1}{3}\left[ \hat{N}_a^2+\hat{N}_b^2 + \hat{N}_a \hat{N}_b +3 \left(\hat{N}_a+\hat{N}_b\right) \right]- \hat{\mathcal{O}}^\dagger \hat{\mathcal{O}} ,\\  \label{cubic__casimir_12}
&&\hspace*{-.9cm}\hat{C}_3=  \frac{1}{18} \left( \hat{N}_a- \hat{N}_b\right)\left(2\hat{N}_a+\hat{N}_b+3\right) \left(\hat{N}_a+2\hat{N}_b+3\right)+  \frac{ \hat{N}_a-\hat{N}_b}{2} \hat{\mathcal{O}}^\dagger \hat{\mathcal{O}}.
\end{eqnarray}

The Hilbert space $\mathcal{F}$ can be split into the subspaces $\mathcal{F}_{n_a, n_b}$:
\begin{equation}
 \mathcal{F}_{n_a, n_b} \equiv span \left\{ |\cdot\,\rangle: \sum_\alpha n_a^\alpha=n_a, \, \sum_\alpha n_b^\alpha=n_b  \right\}, \quad\quad \mathcal{F}=\oplus\mathcal{F}_{n_a, n_b},
\end{equation}
where the direct sum $\oplus$ is over $n_a$ and $n_b$.
In $\tilde{\mathcal{F}}_{n_a, n_b}\equiv    \mathcal{F}_{n_a, n_b}\cap \ker(\hat{\mathcal{O}})$, the Casimirs take fixed values: 
\begin{eqnarray}
&&\hat{C}_2\Big|_{\tilde{\mathcal{F}}_{n_a, n_b}}=c_2 \mathbb{I}, \quad\quad \hat{C}_3\Big|_{\tilde{\mathcal{F}}_{n_a, n_b}}=c_3\mathbb{I},
\end{eqnarray}
with
\begin{eqnarray}
c_2 &=& \frac{1}{3}\left[ n_b^2+n_a^2 +n_b n_a +3 (n_a+n_b) \right]=\textrm{fixed},\\
 c_3 &=& \frac{1}{18} \left( n_a- n_b\right)\left(2n_a+n_b+3\right) \left(n_a+2n_b+3\right) = \textrm{fixed}.
\end{eqnarray}
Also in $\tilde{\mathcal{F}}_{n_a, n_b}$,
\begin{equation}
\hat{s}_i \hat{s}_i = \textrm{fixed}, \quad\quad d_{ijk}\hat{s}_i \hat{s}_j \hat{s}_k=\textrm{fixed}.
\end{equation}
Therefore, the algebra of $\hat{s}_i$'s restricted to $\tilde{\mathcal{F}}_{n_a,n_b}$ describes a 6-dimensional fuzzy space $\mathcal{M}_F^6$.

Each $\tilde{\mathcal{F}}_{n_a,n_b}$ is a carrier space of a finite dimensional irrep of $SU(3)$. This representation is characterized by a pair of positive integers $(p,q)=(n_a, n_b)$\footnote{The Casimirs of a $(p,q)$ representation of $SU(3)$ are $$\hat{C}_2 = \frac{1}{3}[p^2+q^2+pq +3 (p+q)]\,\mathbb{I},\quad \hat{C}_3 = \frac{1}{18}(p-q) (2p+q+3)(p+2q+3)\mathbb{I}.$$} and is of  dimension 
\begin{equation}
dim_{SU(3)} = \frac{1}{2} \left(n_b+n_a+2\right) \left(n_b+1\right) \left(n_a+1\right). 
\end{equation}
 The $\hat{y}_i$'s (and $\hat{s}_i$'s) are square matrices in $\tilde{\mathcal{F}}_{n_a,n_b}$.
 Thus $\mathcal{M}_F^6$ is the fuzzy version of ${SU(3)}/({U(1)\times U(1)})$ and  (\ref{map1}) is a noncommutative $U(1)\times U(1)$ fibration.

Note that in the above construction neither $n_a$ nor $n_b$ can be chosen to be zero. In case the construction is done with only one set of oscillators (either $\hat{a}_\alpha$'s or $\hat{b}_\alpha$'s), one  would get fuzzy $\mathbb{C} P^2$, as in \cite{Trivedi:2000mq, Alexanian:2001qj}. Nevertheless, for $n_a >> n_b$ (or $n_b >> n_a$),  $\mathcal{M}_F^6$ looks like fuzzy $\mathbb{C}P^2$ in some sense  \cite{Trivedi:2000mq}.

\section{Noncommutative line bundle}\label{sec_3}

Let $\mathcal{H}_{n_a,n_b \rightarrow l_a,l_b}$ be the space of linear operators $\Phi$, which map $ \tilde{ \mathcal{F}}_{n_a,n_b}$ to $ \tilde{\mathcal{F}}_{l_a,l_b}$:
\begin{equation}
\Phi: \tilde{ \mathcal{F}}_{n_a,n_b}\rightarrow \tilde{\mathcal{F}}_{l_a,l_b}, \quad \quad \Phi\in\mathcal{H}_{n_a,n_b\rightarrow l_a, l_b}.
\end{equation}
In general, these $\Phi$'s are rectangular matrices.

 $\mathcal{H}_{n_a,n_b \rightarrow n_a,n_b}$ is a noncommutative algebra  which maps $\tilde{ \mathcal{F}}_{n_a,n_b}\rightarrow \tilde{\mathcal{F}}_{n_a,n_b}$ and any $\Phi \in \mathcal{H}_{n_a,n_b \rightarrow n_a,n_b}$ is a square matrix.  In this algebra, the rotations are generated by the adjoint action of $\hat{y}_i^{(n_a,n_b)}$ ($\hat{y}_i^{(n_a,n_b)}$ is the restriction of $\hat{y}_i$ in $\tilde{\mathcal{F}}_{n_a, n_b}$):
\begin{eqnarray}
Ad\left(\hat{y}^{(n_a, n_b)}_i\right)\Phi\equiv\hat{\mathcal{L}}_i \Phi \equiv [\hat{y}_i^{(n_a,n_b)}, \Phi], \quad \quad \Phi \in \mathcal{H}_{n_a,n_b \rightarrow n_a,n_b}.
\end{eqnarray}
$\hat{\mathcal{L}}_i$'s generate a $SU(3)$:
\begin{equation}\label{rotation1}
[\hat{\mathcal{L}}_i,\hat{\mathcal{L}}_j]=if_{ijk}\hat{\mathcal{L}}_k.
\end{equation}

When  $n_a\neq l_a, n_b\neq l_b$, the spaces $ \mathcal{H}_{n_a,n_b \rightarrow l_a,l_b}$  are noncommutative bimodules and any $\Phi \in\mathcal{H}_{n_a,n_b \rightarrow l_a,l_b}$ is a rectangular matrix. For the bimodules, the generators of the $SU(3)$ in (\ref{rotation1}) act by a left- and a right-multiplication: 
\begin{equation}
\hat{\mathcal{L}}_i \Phi = \hat{y}_i^{(l_a,l_b)}\Phi-\Phi \hat{y}_i^{(n_a,n_b)}.
\end{equation}
This $SU(3)$ action is  reducible   and we will give its explicit decomposition shortly.
Any element  $\Phi \in\mathcal{H}_{n_a,n_b \rightarrow l_a,l_b}$ can be expanded in the basis of the eigenvectors of $\hat{\mathcal{L}}_3$, $\hat{\mathcal{L}}_8$, $\hat{\mathcal{L}}_i\hat{\mathcal{L}}_i$ and $d_{ijk}\hat{\mathcal{L}}_i\hat{\mathcal{L}}_j\hat{\mathcal{L}}_k$.

To construct the basis vectors, let us start as follows. 
The operator 
\begin{equation}
\hat{f} = (\hat{a}_3^\dagger)^{\tilde{l}_a}\left(\hat{a}_2\right)^{\tilde{n}_a} (\hat{b}_2^\dagger)^{\tilde{l}_b}(\hat{b}_3)^{\tilde{n}_b}
\end{equation}
is an element  of $\mathcal{H}_{n_a,n_b \rightarrow l_a,l_b}$ if $(\tilde{l}_a, \tilde{n}_a, \tilde{l}_b, \tilde{n}_b)$ are positive  integers satisfying 
\begin{eqnarray}\begin{array}{lll}\label{kakab}
\kappa_a \equiv l_a - n_a=\tilde{l}_a - \tilde{n}_a, \quad\quad \kappa_b\equiv  l_b - n_b = \tilde{l}_b - \tilde{n}_b.
\end{array}
\end{eqnarray}
It is easy to see that
\begin{eqnarray}\begin{array}{lll}
\hat{U}_+ \hat{f} \equiv \left(\hat{\mathcal{L}}_1 + i \hat{\mathcal{L}}_2\right) \hat{f}= 0, \\
\hat{V}_+ \hat{f} \equiv \left(\hat{\mathcal{L}}_4 - i \hat{\mathcal{L}}_5\right) \hat{f}= 0, \\
\hat{W}_+ \hat{f} \equiv \left(\hat{\mathcal{L}}_6 - i \hat{\mathcal{L}}_7\right) \hat{f}= 0,\\ 
\hat{\mathcal{L}}_3 \hat{f} = \frac{1}{2}\left(\tilde{n}_a +\tilde{l}_b\right) \hat{f}, \\
\hat{\mathcal{L}}_8 \hat{f} = -\frac{1}{2\sqrt{3}}\left(2\tilde{l}_a +2\tilde{n}_b+ \tilde{n}_a +\tilde{l}_b\right) \hat{f}.
\end{array}
\end{eqnarray}
So $\hat{f}$ is the highest weight vector of the $SU(3)$ representation characterized by $(p,q)$, with
\begin{equation}
p= \tilde{l}_a +\tilde{l}_b+ \tilde{n}_a +\tilde{n}_b, \quad \quad q= \tilde{l}_a +\tilde{n}_b,\quad\quad p\geq q\geq0.
\end{equation}
The quadratic  and the cubic Casimirs  for this representation take values
\begin{eqnarray}
&&\mathscr{C}_2= \frac{1}{3}\left( p^2+q^2 -pq +3p\right), \\ 
&&\mathscr{C}_3 = \frac{1}{18}(p-2q) (2p-q+3)(q+p+3), 
\end{eqnarray}
while the dimension  is
\begin{equation}
d= \frac{1}{2} (p-q+1)(p+2)(q+1).
\end{equation}

\hspace*{.2cm}The lower weight vectors belonging to the same irrep $(p,q)$ are generated by the action of the lowering operators $\hat{U}_- \left(\equiv\hat{\mathcal{L}}_1 - i \hat{\mathcal{L}}_2\right)$, $\hat{V}_- \left(\equiv\hat{\mathcal{L}}_4 + i \hat{\mathcal{L}}_4\right)$ and $\hat{W}_- \left(\equiv\hat{\mathcal{L}}_6+ i \hat{\mathcal{L}}_7\right)$ on $\hat{f}$.

A generic vector belonging to the $(p,q)$ irrep is labelled by  $m_3$ and $m_8$ -- the $\hat{\mathcal{L}}_3$ and $\hat{\mathcal{L}}_8$ values respectively:
\begin{eqnarray}
\begin{array}{lll}
\hat{\mathcal{L}}_3\Psi ^{m_3,m_8}_{p,q}=m_3\Psi ^{m_3,m_8}_{p,q}, & \quad  \hat{\mathcal{L}}_8\Psi ^{m_3,m_8}_{p,q}=m_8 \Psi ^{m_3,m_8}_{p,q}, \\\\
\hat{\mathcal{L}}_i\hat{\mathcal{L}}_i \Psi ^{m_3,m_8}_{p,q}= \mathscr{C}_2\Psi ^{m_3,m_8}_{p,q}, & \quad d_{ijk}\hat{\mathcal{L}}_i\hat{\mathcal{L}}_j\hat{\mathcal{L}}_k \Psi ^{m_3,m_8}_{p,q}=\mathscr{C}_3 \Psi ^{m_3,m_8}_{p,q}.
\end{array}
\end{eqnarray}\\ \\
In the following, we specify the allowed values of $(p,q)$ (i.e.\,\,which irreps occur).  In (\ref{kakab}), $\kappa_a$ and $\kappa_b$ can be both positive and negative. The ranges of the pairs $(\tilde{l}_a, \tilde{n}_a)$ and$(\tilde{l}_b, \tilde{n}_b)$ are different for each choice of the sign of $\kappa_a$ and $\kappa_b$.
Consequently,  the irreps appearing in such maps also differ. We find the irreps for each case.\\
\mbox{}\\
{\bf Case 1:} $ {\bf   l_a \geq n_a }$  {\bf and} ${\bf l_b \geq n_b}$\\
In this case, both $\kappa_a , \kappa_b \geq 0$. The ranges of $\tilde{n}_a$ and $\tilde{n}_b$ are 
\begin{equation}
0\leq \tilde{n}_a \leq n_a, \quad \quad 0\leq \tilde{n}_b \leq n_b.
\end{equation}
 Therefore the allowed values of $p$ and $q$ are
\begin{table}[H]
\begin{center}
\begin{tabular}{|c|c|c|c|c|c|}
\hline&&&&&\\
$\tilde{n}_a$ &$\tilde{l}_a$& $\tilde{n}_b$ &$\tilde{l}_b$ &$p$& $q$ \\
\hline \hline
0&$\kappa_a$&0& $\kappa_b$& $\kappa_a+\kappa_b$&$\kappa_a$ \\
\hline
1&$\kappa_a+1$&0& $\kappa_b$& $\kappa_a+\kappa_b+2$& $\kappa_a+1$\\
\hline 
0&$\kappa_a$&1& $\kappa_b+1$& $\kappa_a+\kappa_b+2$&$\kappa_a+1$ \\
\hline 
$\cdots$ &$\cdots$ &$\cdots$ &$\cdots$ &$\cdots$&$\cdots$\\
\hline
$n_a$& $l_a$ &$n_b$&$l_b$ &$J_a+J_b$& $ l_a+n_b$\\
\hline
\end{tabular}
\end{center}
\end{table}
\vspace*{-.6cm}
\mbox{}\\
where $J_a\equiv l_a+n_a$ and $J_b\equiv l_b+n_b$.\\
\mbox{}\\
{\bf Case 2:} $ {\bf   l_a \leq n_a }$  {\bf and} ${\bf l_b \geq n_b}$\\
Here, $\kappa_a \leq 0$ and $\kappa_b \geq 0$ and 
\begin{equation}
0\leq \tilde{l}_a \leq l_a, \quad \quad 0\leq \tilde{n}_b \leq n_b.
\end{equation}
Hence  $p$ and $q$ take the following values:
\begin{table}[H]
\begin{center}
\begin{tabular}{|c|c|c|c|c|c|}
\hline&&&&&\\
$\tilde{l}_a$ &$\tilde{n}_a$& $\tilde{n}_b$ &$\tilde{l}_b$ &$p$& $q$ \\
\hline \hline
0&$-\kappa_a$&0& $\kappa_b$& $-\kappa_a+\kappa_b$&$0$ \\
\hline
1&$-\kappa_a+1$&0& $\kappa_b$& $-\kappa_a+\kappa_b+2$& $1$\\
\hline 
0&$-\kappa_a$&1& $\kappa_b+1$& $-\kappa_a+\kappa_b+2$&$1$ \\
\hline 
$\cdots$ &$\cdots$ &$\cdots$ &$\cdots$ &$\cdots$&$\cdots$\\
\hline
$l_a$& $n_a$ &$n_b$&$l_b$ &$J_a+J_b$& $ l_a+n_b$\\
\hline
\end{tabular}
\end{center}
\end{table}
\mbox{}\\
{\bf Case 3:} $ {\bf   l_a \geq n_a }$  {\bf and} ${\bf l_b \leq n_b}$\\
When $\kappa_a \geq 0$ and $\kappa_b \leq 0$, the ranges of $\tilde{n}_a$ and $\tilde{l}_b$ are 
\begin{equation}
0\leq \tilde{n}_a \leq n_a, \quad \quad 0\leq \tilde{l}_b \leq l_b.
\end{equation}
 Then the allowed values of $p$ and $q$ are  
\begin{table}[H]
\begin{center}
\begin{tabular}{|c|c|c|c|c|c|}
\hline&&&&&\\
$\tilde{n}_a$ &$\tilde{l}_a$& $\tilde{l}_b$ &$\tilde{n}_b$ &$p$& $q$ \\
\hline \hline
0&$\kappa_a$&0& $-\kappa_b$& $\kappa_a-\kappa_b$&$\kappa_a-\kappa_b$ \\
\hline
1&$\kappa_a+1$&0& $-\kappa_b$& $\kappa_a-\kappa_b+2$& $\kappa_a-\kappa_b+1$\\
\hline 
0&$\kappa_a$&1& $-\kappa_b+1$& $\kappa_a-\kappa_b+2$&$\kappa_a-\kappa_b+1$ \\
\hline 
$\cdots$ &$\cdots$ &$\cdots$ &$\cdots$ &$\cdots$&$\cdots$\\
\hline
$n_a$& $l_a$ &$l_b$&$n_b$ &$J_a+J_b$& $ l_a+n_b$\\
\hline
\end{tabular}
\end{center}
\end{table}
\mbox{}\\
{\bf Case 4:} $ {\bf   l_a \leq n_a }$  {\bf and} ${\bf l_b \leq n_b}$\\
In this case, $\kappa_a \leq 0$ and $\kappa_b \leq 0$ and 
\begin{equation}
0\leq \tilde{l}_a \leq l_a, \quad \quad 0\leq \tilde{l}_b \leq l_b.
\end{equation}
Thus the irreps have $(p,q)$ values 
\begin{table}[H]
\begin{center}
\begin{tabular}{|c|c|c|c|c|c|}
\hline&&&&&\\
$\tilde{l}_a$ &$\tilde{n}_a$& $\tilde{l}_b$ &$\tilde{n}_b$ &$p$& $q$ \\
\hline \hline
0&$-\kappa_a$&0& $-\kappa_b$& $-(\kappa_a+\kappa_b)$&$-\kappa_b$ \\
\hline
1&$-\kappa_a+1$&0& $-\kappa_b$& $-(\kappa_a+\kappa_b)+2$& $-\kappa_b+1$\\
\hline 
0&$-\kappa_a$&1& $-\kappa_b+1$& $-(\kappa_a+\kappa_b)+2$&$-\kappa_b+1$ \\
\hline 
$\cdots$ &$\cdots$ &$\cdots$ &$\cdots$ &$\cdots$&$\cdots$\\
\hline
$l_a$& $n_a$ &$l_b$&$n_b$ &$J_a+J_b$& $ l_a+n_b$\\
\hline
\end{tabular}
\end{center}
\end{table}

Any arbitrary element $\Phi\in\mathcal{H}_{n_a,n_b\rightarrow l_a,l_b}$ can be expressed in terms of the $SU(3)$ harmonics as 
\begin{equation}\label{phi_1}
\Phi =\sum_{m_3,m_8,p,q} c ^{m_3,m_8}_{p,q}\Psi ^{m_3,m_8}_{p,q}, \quad \quad c^{m_3,m_8}_{p,q}\in \mathbb{C}.
\end{equation}
These $\Phi$'s are identified as the noncommutative analogue of the sections of the associated line bundle.

\subsubsection*{Topological charge}
The sections of the associated line bundle carry two topological charges,  corresponding to each $U(1)$ fibre.
 In $\mathcal{H}_{n_a,n_b\rightarrow l_a,l_b}$,  we can define two topological charge operators:
\begin{eqnarray}
\hat{K}_a \equiv \frac{1}{2} \left[ \hat{N}_a, \quad \right], \quad\quad \hat{K}_b \equiv \frac{1}{2} \left[ \hat{N}_b,\quad \right].
\end{eqnarray}
It is easy to see that   $\Phi $ in (\ref{phi_1})   has topological charges $(\kappa_a,\kappa_b)$  given by
\begin{equation}
\hat{K}_a \Phi = \frac{\tilde{l}_a -\tilde{n}_a}{2} \Phi= \frac{\kappa_a}{2} \Phi, \quad\quad \hat{K}_b \Phi = \frac{\tilde{l}_b -\tilde{n}_b}{2} \Phi=\frac{\kappa_b}{2} \Phi,\quad\quad \kappa_a, \kappa_b \in \mathbb{Z}.
\end{equation}
Therefore $\Phi$ is a section of a complex line bundle with topological charges $(\kappa_a, \kappa_b)$.

\section{The Dirac operator}\label{sec_4}

${SU(3)}/({U(1)\times U(1)})$ is a 6-dimensional space embedded in $\mathbb{R}^8$. This space is curved and
non-symmetric. Also, as $U(1) \times U(1) \subset Spin(6) \cong SU(4)$, this space admits a spin structure.
In the commutative case, the Dirac operator contains three terms: the kinetic term, the spin-connection
term and the background monopole term (if any)~\cite{Dolan:2003bj}.

There are many possible ways to obtain the Dirac operator on this space (for example, \cite{Balachandran:2002bj}).
In the fuzzy case, we do so by constructing a Ginsparg-Wilson algebra on $\mathcal{M}_F^6$.

We look at the zero charge sector first.
On $\mathcal{H}_{n_a,n_b\rightarrow n_a,n_b}$,  $\hat{y}_i$ has left and right actions
\begin{equation}
\hat{y}_i^Lf\equiv\hat{y}_if, \quad \quad \hat{y}_i^Rf\equiv f\hat{y}_i.
\end{equation}
These satisfy (\ref{algebra_12})-(\ref{cubic__casimir_12}):
\begin{eqnarray}
\begin{array}{cccc}
 & [\hat{y}_i^L,\hat{y}_j^L]=if_{ijk}\hat{y}_k^L, \quad \quad [\hat{y}_i^R,\hat{y}_j^R]=-if_{ijk}\hat{y}_k^R,
\quad \quad [\hat{y}_i^L,\hat{y}_j^R]=0, \\ \\
& \hat{y}_i\hat{y}_i\equiv \hat{y}_i^L\hat{y}_i^L=\hat{y}_i^R\hat{y}_i^R=c_2 \mathbb{I}, \quad \quad d_{ijk}\hat{y}_i^L\hat{y}_j^L\hat{y}_k^L=
d_{ijk}\hat{y}_i^R\hat{y}_j^R\hat{y}_k^R=c_3 \mathbb{I}. 
\end{array}
\end{eqnarray}

The $\gamma$-matrices on $\mathbb{R}^8$ are $16\times 16$ matrices which generate the Clifford algebra
\begin{equation}
\{\gamma_i,\gamma_i\}=2\delta_{ij}, \quad \quad \gamma_i^\dagger=\gamma_i, \quad \quad i=1,2,\ldots,8.
\end{equation}
Using $\gamma_i$'s we can construct
\begin{equation}
t_i\equiv\frac{1}{4i}f_{ijk}\gamma_j\gamma_k, \quad \quad [t_i,\gamma_j]=if_{ijk}\gamma_k,
\end{equation}
which generate a $SU(3)$:
\begin{equation}
[t_i,t_j]=if_{ijk}t_k.
\end{equation}
We can define
\begin{equation}
\Gamma\equiv\frac{1}{a}\gamma_i(\hat{y}_i^L+\frac{1}{3}t_i), \quad \quad
\tilde{\Gamma}\equiv-\frac{1}{a}\gamma_i(\hat{y}_i^R-\frac{1}{3}t_i),
\end{equation}
where the normalization $a$ is given by 
\begin{equation}
a^2 \mathbb{I}=\hat{y}_i\hat{y}_i+\frac{1}{3}t_it_i, \quad\quad t_it_i=3 \mathbb{I}, \quad\quad \hat{y}_i\hat{y}_i = c_2 \mathbb{I}.
\end{equation} 
$\Gamma$ and $\tilde{\Gamma}$ generate of a Ginsparg-Wilson algebra:
\begin{equation}
\mathcal{A}_{GW}=\{\Gamma,\tilde{\Gamma}: \,\,\, \Gamma^2=\mathbb{I}=\tilde{\Gamma}^2, \,\, \Gamma^\dagger=\Gamma, \,\,
\tilde{\Gamma}^\dagger=\tilde{\Gamma}\}.
\end{equation}

From this algebra, one can construct  a Dirac operator $\mathcal{D}$ 
\begin{eqnarray}\label{dirac_defn}
\mathcal{D}=a\left(\Gamma+\tilde{\Gamma}\right)=\gamma_i\hat{\mathcal{L}}_i+\frac{2}{3}\gamma_it_i, \quad\quad \textrm{where \,\,}\hat{\mathcal{L}}_i = \hat{y}_i^{L} +\hat{y}_i^{R},
\end{eqnarray}
and a chirality operator $\Gamma_{ch}$
\begin{eqnarray}
 &&\Gamma_{ch}=a\left(\Gamma-\tilde{\Gamma}\right) = \gamma_i \left(\hat{y}_i^{L} +\hat{y}_i^R\right).
\end{eqnarray}
 It is easy to check that 
\begin{eqnarray}
&& \{\mathcal{D},\Gamma_{ch}\}=0,\\
&&\nu\equiv \textrm{index}_{\mathcal{D}}=\textrm{Tr}\,(\Gamma_{ch}) \quad\quad \textrm{(index theorem)}.
\end{eqnarray}
It is interesting to note that in (\ref{dirac_defn}),  the second term ($\sim \gamma_i t_i$)  is the spin-connection term in~\cite{Dolan:2003bj}.

The Dirac operator is of the form 
\begin{eqnarray}
\mathcal{D}= \left(\begin{array}{cccc}
0 & A\\
A^\dagger &0
\end{array}\right).
\end{eqnarray}
When the action of  $\mathcal{D}$ is 
restricted to the algebra $\left(\mathcal{H}_{n_a,n_b\rightarrow n_a,n_b}\times Mat_{16}\right)$, it is the Dirac operator on $\mathcal{M}_F^6$.
In this case, there is no monopole background, and hence the gauge field contribution to the Dirac operator is zero.

On the bimodule $\mathcal{H}_{n_a,n_b\rightarrow l_a,l_b}$ with $n_a\neq l_a, \, n_b\neq l_b$ or
either, there is a background monopole. On this bimodule, $\hat{\mathcal{L}}_i$ is the covariant derivative which includes the monopole contribution.
Hence, if we restrict $\mathcal{D}$ to the bimodule $\left(\mathcal{H}_{n_a,n_b\rightarrow l_a,l_b}\times Mat_{16} \right)$, we
automatically incorporate the background monopole information. There is no need to add the monopole term in the Dirac operator.
Rather, restricting the algebra of $\mathcal{D}$ to the proper subspace  accounts for monopoles. 

\subsection{Zero modes of the Dirac operator}
${SU(3)}/({U(1) \times U(1)})$ is a non-symmetric space with positive curvature. On this space, there is an additional connection due to the torsion, which appears in  the square of the Dirac operator \cite{Dolan:2003bj}:
\begin{equation}
\mathcal{D}^2 = -\nabla^2 + \textrm{curvature} + \textrm{torsion}+ \textrm{possible gauge field contribution}.
\end{equation}
The Dirac Laplacian $\nabla^2$ on this coset space is a positive operator. For the Dirac operator to have zero modes, we would need the cancellation of the lowest eigenvalue of the Laplacian with the lowest eigenvalue of the sum of the curvature, torsion and gauge field contributions. 
These considerations require that   the number of zero modes of the Dirac operator on ${SU(3)}/({U(1)\times U(1)})$ is given by the dimension of the $SU(3)$ irrep with the minimum value of the quadratic Casimir $\mathscr{C}_2$ \cite{Dolan:2003bj}.
We adopt the same requirement to compute the number of zero modes   in the fuzzy case. \\
\mbox{}\\ 
{\bf Case 1:} $ {\bf  \kappa_a \geq 0}$  {\bf and} ${\bf \kappa_b \geq 0}$\\
In this case  $ \mathscr{C}_2$ takes
the minimum value in the representation with  $(p,q)=(\kappa_a+\kappa_b,\, \kappa_a )$:
\begin{equation}
\mathscr{C}_2^{\textrm{min}}=\frac{1}{3}\left(\kappa_a^2+\kappa_b^2+  \kappa_a \kappa_b\right)+\kappa_a+\kappa_b.
\end{equation}
The dimension of this representation is the index of $\mathcal{D}$:
\begin{equation}
\nu= d^{\textrm{min}}=\frac{1}{2}(\kappa_a+\kappa_b+2)(\kappa_a+1)(\kappa_b+1).
\end{equation}\\
The discussion for the other cases is similar:\\ \\
{\bf Case 2:} $ {\bf   \kappa_a \leq 0 }$  {\bf and} ${\bf \kappa_b \geq 0}$
\begin{eqnarray}\left.
\begin{array}{llll}
(p,q)=(-\kappa_a+\kappa_b,\,0), \\ \\
\mathscr{C}_2^{\textrm{min}}=\frac{1}{3}\left(-\kappa_a+\kappa_b\right)^2 -\kappa_a+\kappa_b, \\ \\
\nu= d^{\textrm{min}}=\frac{1}{2}(-\kappa_a+\kappa_b+2)(-\kappa_a+\kappa_b+1).
\end{array} \right.
\end{eqnarray}
\\
{\bf Case 3:} $ {\bf  \kappa_a \geq 0 }$  {\bf and} ${\bf \kappa_b \leq 0}$\\
 \begin{eqnarray}\left.
\begin{array}{llll}
(p,q)=(\kappa_a-\kappa_b,\,\kappa_a-\kappa_b), \\ \\
\mathscr{C}_2^{\textrm{min}}=\frac{1}{3}\left(\kappa_a-\kappa_b\right)^2+ \kappa_a-\kappa_b, \\ \\
\nu= d^{\textrm{min}}=\frac{1}{2}(\kappa_a-\kappa_b+2)(\kappa_a-\kappa_b+1).
\end{array} \right.
\end{eqnarray}
\\
{\bf Case 4:} $ {\bf  \kappa_a \leq 0 }$  {\bf and} ${\bf \kappa_b \leq 0}$\\ 
\begin{eqnarray}\left.
\begin{array}{llll}
(p,q)=(-\kappa_a-\kappa_b,\,-\kappa_b), \\ \\
\mathscr{C}_2^{\textrm{min}}=\frac{1}{3}\left(\kappa_a^2+\kappa_b^2 + \kappa_a \kappa_b \right)-\kappa_a-\kappa_b, \\ \\
\nu= d^{\textrm{min}}=\frac{1}{2}(-\kappa_a-\kappa_b+2)(-\kappa_a+1)(-\kappa_b+1).
\end{array} \right.
\end{eqnarray}
\\

When there is no monopole, $\kappa_a=0=\kappa_b$ and  the index is
\begin{equation}
\nu=1,
\end{equation}
which is consistent with~\cite{Dolan:2003bj,Alexanian:2001qj}.
Also, as in the commutative case, the index $\nu$ gives $\mathrm{Tr}\, (F\wedge F\wedge F)$ for the monopole fields.

\section{Conclusions}

Our realization of $\mathcal{M}_F^6$ can be used to study large $N$ limits of  matrix models of $D$-branes. Among other things, $\mathcal{M}_F^6$  as the vacua of the matrix model   (\ref{matrix_action}) can be in an irreducible or reducible representation. 
It is easy to generalize the Schwinger construction of $\mathcal{M}_F^6$ by using Brandt-Greenberg oscillators and obtain  reducible algebras of $\mathcal{M}_F^6$, as in \cite{Acharyya:2014nfa}. We can obtain the quantum states for those reducible $\mathcal{M}_F^6$'s using the prescription of Gelfand-Naimark-Segal (GNS). Those quantum states will be inherently mixed and will carry entropy, which is typically large. These informations will  play a vital role in the understanding of the vacua and tachyon condensations in the matrix model.

 Using the Dirac operator and the index theory, one may try to construct the spinor bundle on $\mathcal{M}_F^6$ and thus find the supersymmetric analogue of $X_F^8 \rightarrow \mathcal{M}_F^6$. The  super conifold in the continuum has various interesting features \cite{Ricci:2005cp} which  should  be manifest  in the fuzzy version as well. We leave this for a future investigation.

The fermion-doubling  problem on this finite dimensional space can also be studied. It has been shown that there is no fermion-doubling on the fuzzy sphere \cite{Balachandran:1999qu, Balachandran:2003ay}. There might be such dramatic   consequences on $\mathcal{M}_F^6$ as well.

\mbox{}\\ \\
\textbf{Acknowledgements}\\
V.E.D. thanks I.I.Sc. for hospitality during her stay from September 2014 to October 2015.
We would like to thank A.~P.~Balachandran, Sachindeo Vaidya and Keshav Dasgupta for illuminating discussions and suggestions. We are indebted to Diptiman Sen, who pointed out an error in an earlier version of the draft.

\end{document}